\documentstyle[aps,prl,twocolumn,epsf]{revtex}
\begin{document}

\title{Decoherence via Dynamical Casimir Effect}

\author{Diego A.\ R.\ Dalvit $^{1}$ and
Paulo A. Maia Neto $^{2}$}

\address{$^1$ \it T-6, Theoretical Division, MS B288, Los Alamos National
Laboratory, Los Alamos, NM 87545, USA}

\address{$^2$
\it Instituto de F\'{\i}sica, UFRJ, Caixa Postal 68528, 21945-970 Rio
de Janeiro, Brazil}

\date{\today}

\maketitle

\begin{abstract}
We derive a master equation for a mirror interacting with the vacuum 
field via radiation pressure. The dynamical Casimir effect leads to 
decoherence of a `Schrodinger cat' state in a time scale
that depends on the degree of 
`macroscopicity' of the state components, and which
may be
 much shorter
than the relaxation time scale.
Coherent states are selected by the interaction as pointer states.

\end{abstract}

Within the framework of Quantum Mechanics, a closed system 
may be found in any quantum state of the Hilbert space. As pointed out
by Schr\"odinger~\cite{Schr}, this is in apparent contradiction with the
classical behavior of macroscopic systems.
However, macroscopic systems are seldom isolated, and the interaction with 
the environment engenders the decay of most states into a statistical mixture
 of `pointer states,' which are linked to classical properties of the 
system~\cite{physicstoday}.
Coherent superpositions of pointer states
decohere into a statistical mixture in a time scale which is usually
of the order of the damping time divided by some parameter representing the
degree of `classicality' of the states.
The decoherence time scale for a microwave field in a high-Q
superconducting cavity was recently measured~\cite{haroche} to be
in agreement with such prediction~\cite{luiz}.

Several different heuristic models for the coupling with the environment
have been considered~\cite{all}. In this letter, we show that the coupling
with the quantum vacuum field via radiation pressure provides
a more fundamental, {\it ab initio} model for decoherence.
The Casimir effect for moving boundaries has attracted a lot of
interest recently~\cite{all-Casimir}. The vacuum radiation pressure
force dissipates the mechanical energy of an oscillating mirror, and the 
associated
photon emission effect could in principle be measured 
experimentally~\cite{astrid}. Usually, one assumes that the  mirror 
follows a prescribed trajectory, thus neglecting the recoil effect.
However, here we want to focus on the mirror
as a dynamical quantum system,  hence the need to take
the full mirror-plus-field dynamics into account.
Jaekel and Reynaud treated this problem by using
linear response theory~\cite{jr}, in order to calculate the fluctuations 
of the position of a dispersive
 mirror driven by the vacuum radiation pressure. Mass corrections were 
also obtained in Refs.~\cite{jr2} and \cite{barton}.

In this letter, we consider
a nonrelativistic
partially-reflecting mirror of mass $M$ (position $q$ and momentum
$p$) in the harmonic potential of frequency $\omega_0,$  and under the 
action of vacuum radiation
pressure. We take a scalar field  in $1+1$ dimensions, and neglect
third and higher order terms in $v/c,$ where $v$ is the mirror's velocity
(we set $c=1$). We start from the Hamiltonian formalism developed in 
Ref.~\cite{barton}.
The Hamiltonian is given by
$H = H_M + H_F +  H_{\rm int},$
where
\begin{equation}
H_M = {p^2\over 2M}  + {M\omega_0^2\over 2} q^2, \label{h2}
\end{equation}
\begin{equation}
H_F = \int {dx\over 2} \left[\Pi^2 + (\partial_x\phi)^2\right] +
 \Omega \phi^2(x=0) \label{h3}
\end{equation}
is the Hamiltonian for the scalar field $\phi$
($\Pi= \partial_t \phi$ is its momentum canonically conjugate)
under the boundary condition
corresponding to a partially-reflecting mirror at rest at $x=0,$
where the coupling constant $\Omega$ also plays the role of a
transparency frequency, the
frequency-dependent reflection amplitude being 
$R(\omega)=-i\Omega/(\omega + i \Omega)$~\cite{jr2}\cite{barton}.
Since the emitted photons have frequencies smaller than $\omega_0,$
the perfectly-reflecting limit corresponds to $\omega_0\ll \Omega.$
We allow in principle
for arbitrary values of $\omega_0/\Omega,$ but assume from the
start that $\hbar \omega_0/M,$ which is of the order of
the recoil velocity of the mirror, is very small.
The interaction Hamiltonian $ H_{\rm int}$ describes, on one hand,
the modification of the boundary condition for the field due to the motion 
of the mirror, and, on the other hand, the modification of the mirror's 
motion engendered by the radiation pressure force. The first effect leads to
the emission of photon pairs out of the vacuum state (dynamical Casimir
effect), whereas the second leads to dissipation and decoherence
of the mirror's motion, as
shown below. To second order in $v/c,$ we have
\begin{equation}
 H_{\rm int} = -{p {\cal P}\over M} + {{\cal P}^2\over 2 M}
-{1\over 2} \Omega \phi^2(0) {p^2\over M^2},\label{h4}
\end{equation}
where
${\cal P} = -\int dx \partial_x \phi \, \partial_t \phi $
is the field momentum operator. In the r.-h.-s. of~(\ref{h4}), the
first term is the most important, yielding the effects of dissipation and
decoherence. The second term does not depend on the mirror's variables,
and hence
will be of no relevance here, whereas the third term, being already of
second order in $v/c$, is taken only to first order in perturbation theory. 
As discussed
in Ref.~\cite{barton}, it provides a contribution to the mirror's 
mass shift.

We derive a master equation for the reduced density matrix of the mirror
$\rho(t)$ by assuming that at $t=0$
the mirror and the field are not correlated, so that the density matrix of the
combined system ${\tilde \rho}$ is written as  ${\tilde \rho}(0)=\rho(0)\otimes
\rho_F,$ where $\rho_F$ is the density matrix of the field alone.
Then we compute ${\tilde \rho}(t)$ up to second order in the perturbation Hamiltonian $H_{\rm int}.$ Note that the small perturbation parameter
is the mirror's velocity $v/c,$ and not
the coupling constant $\Omega,$ which is
incorporated in the field Hamiltonian $H_F.$
In fact, one may diagonalize $H_M+H_F$ for arbitrary values of
$\Omega,$ writing field eigenfunctions containing reflection
and transmission coefficients so as to satisfy the boundary
conditions at $x=0$~\cite{barton}.
Finally, the master equation for $\rho(t)$ is obtained by tracing
${\tilde \rho}(t)$ over the field variables, taking the
field to be in the vacuum state. 
We find
\begin{eqnarray}
i \hbar \dot\rho&=& [H_M - \frac{\Delta M(t)}{M} \frac{p^2}{2 M},\rho] -
\Gamma(t) [p,\{q,\rho\}] \nonumber \\
& & - \frac{i}{\hbar} D_1(t) [p,[p,\rho]] - \frac{i}{\hbar} 
D_2(t) [p,[q,\rho]].
\label{master}
\end{eqnarray}
The mass shift in~(\ref{master}) is given by
$\Delta M(t)=\Delta M_1+\Delta M_2(t),$ where the cut-off
dependent $\Delta M_1 = <\Omega \phi^2(0)>$ is
the
only (first order) contribution of the $p^2$  term in Eq.~(\ref{h4}). 
It was derived earlier by different methods in ~\cite{jr2} and 
\cite{barton}.
Except for $\Delta M_1,$ 
the terms 
in~(\ref{master}) come from 
second order perturbation theory. The corresponding coefficients 
are calculated from vacuum correlation functions of the momentum 
operator. 
The mass shift $\Delta M_2(t)$ and the
damping coefficient $\Gamma(t)$ are obtained from the anti-symmetric
correlation function $\xi(t)=\langle [{\cal P}(t),{\cal P}(0)] \rangle,$
 which is connected to the susceptibility 
describing how the field momentum is affected by 
the motion of the
mirror (and the corresponding modification of the boundary conditions).
In fact, we show below that $\Gamma(t)$ is closely connected to the
photon emission effect and the associated radiation reaction force 
that damps the motion so as to enforce energy conservation.
The diffusion coefficients $D_1(t)$ and $D_2(t)$ are obtained from
the symmetric correlation function
$\sigma(t)= \langle \{{\cal P}(t),{\cal P}(0)\} \rangle
- 2 \langle {\cal P} \rangle^2$, which represents the vacuum fluctuations. 

Since ${\cal P}$ is quadratic in the field operators,
the correlation functions 
are obtained from the two-photon matrix
elements $\langle 0|{\cal P}(t)|\omega_1,\omega_2 \rangle,$
which are  calculated by using the normal mode expansion 
for the field operator.
The spectral density $\Xi(\omega)$ is defined as the Fourier transform 
of $\xi(t).$ 
For $\omega>0,$ $\Xi(\omega)$ results from the contribution of 
two-photon states with $\omega_1+\omega_2=\omega.$
We find 
$\Xi(\omega) = (2/\pi) \hbar^2 \Omega \zeta(\omega/\Omega)$
with
$\zeta(u)=\ln(1+u^2)/(2u)+(\arctan u)/u^2-1/u,$
whereas the Fourier transform of $\sigma(t)$ is 
$\epsilon(\omega) \Xi(\omega)$ ($\epsilon(\omega)$ is the 
sign function).
The transparency frequency $\Omega$ sets 
a frequency scale for the behavior of $\Xi(\omega).$ Thus, for
$\omega \ll \Omega$ the spectral density is linear 
(`ohmic' environment), whereas for high frequencies it goes to zero
as $\zeta(u)\approx \ln(u)/u,$ due to the mirror's transparency
at frequencies $\omega\gg \Omega.$
We find
\begin{equation}
\Delta M_2(t)={2\hbar \Omega \over
\pi^2}\int_{-\infty}^{\infty}  d\omega \zeta(\omega/\Omega)
{\sin^2\left[ (\omega-\omega_0)t/2\right]\over 
\omega -\omega_0},\label{mass2}
\end{equation}
\begin{equation}
\Gamma(t) = {\hbar  \Omega\omega_0\over 2\pi^2 M}
\int_{-\infty}^{\infty} d\omega \zeta(\omega/\Omega) 
{\sin\left[ (\omega_0-\omega )t\right]\over \omega_0-
\omega},\label{Gamma}
\end{equation}
\begin{equation}
D_1(t) = { \hbar^2 \Omega\over 2\pi^2 M^2} \label{D1}
\int_{-\infty}^{\infty} d\omega \epsilon(\omega) \zeta(\omega/\Omega)
{\sin\left[ (\omega_0-\omega)t\right]\over \omega_0-
\omega},
\end{equation}
\begin{equation}
D_2(t) = {\hbar^2 \omega_0 \Omega\over \pi^2 M}\label{D2}
\int_{-\infty}^{\infty} d\omega \epsilon(\omega)\zeta(\omega/\Omega)
{\sin^2\left[ (\omega_0-\omega)t/2\right]\over \omega_0-
\omega}.
\end{equation}

The function $\sin\left[(\omega_0-\omega)t\right]/(\omega_0-\omega)$
in equations~(\ref{Gamma}) and (\ref{D1}) has a peak of
width $2\pi/t$ at $\omega=\omega_0.$
For large times, $\Omega t \gg 1,$ the spectral density
 is approximately constant
over the width of this peak, and then may be taken 
out of the integral, yielding
\begin{equation}
\Gamma = {\hbar \Omega\omega_0\over 2\pi M}
 \zeta(\omega_0/\Omega)\approx{\hbar \omega_0^2\over 12\pi M},\label{gammaf}
\end{equation}
the last approximation being valid in the perfectly-reflecting limit.
If we also assume that $\omega_0 t\gg 1,$ Eq.~(\ref{D1}) yields
$D_1 = \hbar \Gamma/  (M\omega_0). $
Accordingly, for large times the damping and diffusion coefficients
have constant values that result from the contribution of
two-photon states $|\omega_1,\omega_2 \rangle$ such that $\omega_1+\omega_2
=\omega=\omega_0.$ This is precisely the condition satisfied by the
photon pairs generated in the dynamical Casimir effect~\cite{all-Casimir}.
In fact, the damping rate $\Gamma$ as given by Eq.~(\ref{gammaf}) is
directly connected to the dissipative force on the moving mirror
$F= \hbar x^{'''}/ (6 \pi)$ ~\cite{ford}
(for simplicity we consider the perfectly reflecting limit). Indeed, 
the equation of motion for the average position then reads
$
x^{''}= -\omega_0^2 x + \hbar x^{'''}/(6 \pi M),
$
whose solution  in the limit $\hbar \omega_0/M\ll 1$ decays
as $\exp[-\hbar \omega_0^2 t/(12\pi M)]$
in agreement with Eq.~(\ref{gammaf}).

The asymptotic values of the dispersive terms 
$\Delta M_2(t)$ and $D_2(t)$ do not originate, on the other hand, from 
the neighborhood of
$\omega=\omega_0.$ In the perfectly-reflecting limit, we neglect
$\omega_0$ in the denominator in Eq.~(\ref{mass2}), and, when
$\Omega t \gg 1,$
replace the sine squared by one-half. Integration of the resulting
expression over the whole frequency interval yields
$\Delta M_2\approx \hbar\Omega/(2\pi).$
Accordingly, for large times we find the same
mass correction obtained in ~\cite{barton} from
stationary perturbation theory.

From these results, we may address two
fundamental issues: (i) find out  the pointer states; (ii) estimate the
decoherence time scale. In the context considered here, 
pointer states are the most robust
elements of the Hilbert space with respect to the motional interaction with
the vacuum field. A simple test was proposed in Ref.~\cite{Paz}, based on
the idea that for pointer states the rate of information loss is minimum.
Such rate is measured with the help of the linear entropy $ s\equiv
1-{\rm Tr}\rho^2$ ($ s=0$ for a pure state and greater than zero
for a mixture).
We calculate the rate of entropy increase from 
the master equation~(\ref{master}), assuming that the initial state 
is pure:
\begin{equation}
{\dot s}(t) = 2 \Gamma(t) (s(t)-1)  +
\frac{4 D_1(t)}{\hbar^2} (\Delta p)^2 + 
\frac{2 D_2(t)}{\hbar^2} \sigma_{q,p}
\label{s1}
\end{equation}
where $(\Delta p)^2$ is the momentum dispersion and 
$\sigma_{q,p}\equiv \langle \{q,p\} \rangle -2 \langle p \rangle 
\langle q \rangle$
(with all operators evaluated at the same time $t$). 
The first term in Eq.~(\ref{s1}) leads to a decrease of entropy
(hence damping tries to localize the state competing against
diffusion) which does not depend on the
initial state. Thus, it is not relevant for the determination of
the pointer states, and will be left out of our discussion.

\begin{figure}
\centering \leavevmode
\epsfxsize=7cm
\epsfbox{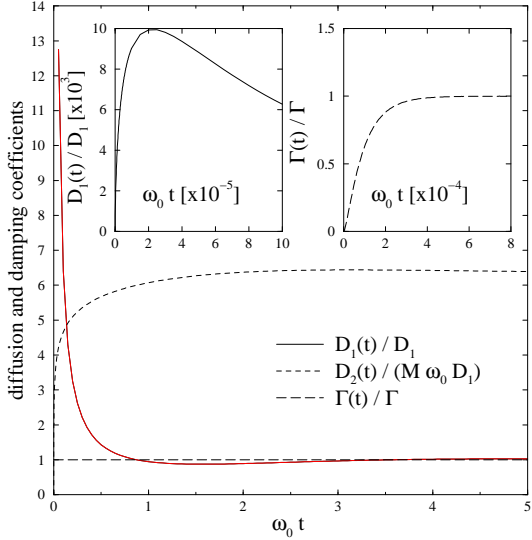}
\caption{\small{Diffusion and damping coefficients as a function of time
in the perfect-reflector limit $\omega_0/\Omega=10^{-4} \ll 1$. Here 
$D_1= \hbar^2 \omega_0/ 12 \pi M^2$ 
and $\Gamma=\hbar \omega_0^2/12 \pi M$
are the asymptotic limits of $D_1(t)$ and $\Gamma(t)$. The insets
show the behavior of these two time-dependent coefficients for short
times.}}
\end{figure}

We first consider
the effect of the last two terms in Eq.~(\ref{s1}) 
in the perfectly-reflecting limit. 
In Fig.~1 we plot the diffusion and damping
coefficients as functions of $\omega_0 t$ for $\omega_0/\Omega=10^{-4}.$ 
$D_1(t)$ develops an initial jolt for times
of the order of $\Omega^{-1}$ and then decreases to the asymptotic value
$(D_1)_{\rm perf}=\hbar^2 \omega_0/(12 \pi M^2)$ 
for $t \sim 1/\omega_0.$
If we integrate Eq.~(\ref{s1}) over many 
periods of oscillation, from $t=0$ to $t=T=n 2 \pi/\omega_0$, 
the contribution to the entropy of the initial jolt 
is negligible, allowing us to 
replace the diffusion coefficients by their constant asymptotic
values. 
When computing
$\sigma_{q,p}(t)$ and $(\Delta p)^2(t)$ in Eq.~(\ref{s1}),
we take the free evolution (corresponding to the harmonic oscillator
Hamiltonian $H_M$) of the mirror's operators $q$ and $p$
(weak coupling approximation). We get
\begin{equation}
s(T)= 2 T {D_1\over \hbar^2} \left[ (\Delta p)_0^2 +
(M\omega_0)^2 (\Delta q)_0^2\right], \label{deltas}
\end{equation}
where 
$(\Delta p)_0^2$ and $(\Delta q)_0^2$ represent the dispersions 
for the initial state. 
Note that $D_2(t)$ does not contribute to the time-averaged entropy production.
The minimum $s(T)$ given the constraint
$\Delta q \Delta p \ge \hbar/2$
is for $\Delta q{}^2=\hbar/(2M\omega_0),$ $\Delta p{}^2=M\hbar\omega_0/2.$
Thus, as in the problem of QBM  with 
interaction Hamiltonian linear in the position operator~\cite{Paz},
the pointer basis consists of coherent states. 

\begin{figure}[h]
\centering \leavevmode
\epsfxsize=7cm
\epsfbox{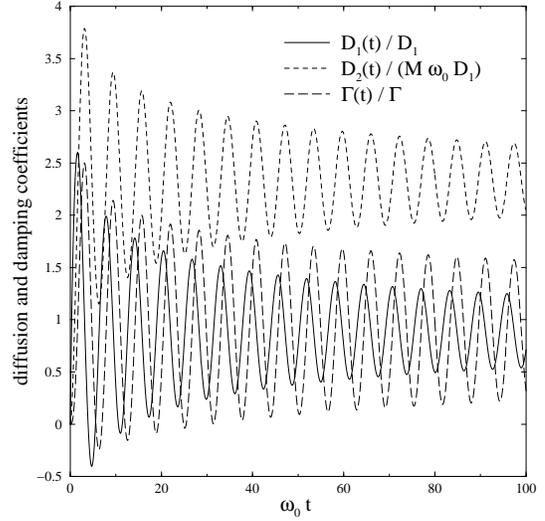}
\caption{\small{Diffusion and damping coefficients as a function of time
in the high-transmission limit $\omega_0/\Omega=10^{4} \gg 1$. Here
$D_1=\hbar^2 \Omega^2 \ln(\omega_0/\Omega) / 2 \pi M^2 \omega_0$ and
$\Gamma=\hbar \Omega^2 \ln(\omega_0/\Omega)/2 \pi M$ are the asymptotic 
limits of $D_1(t)$ and $\Gamma(t)$.}}
\end{figure}

The opposite limit $\omega_0 \gg \Omega$ corresponds to dominant frequencies of
the environment slow with respect to the mirror's own time scale. 
However, since the spectral density $\Xi(\omega)$ decays 
too slowly for $\omega \gg \Omega,$ 
field frequencies of the order of $\omega_0$ provide a significant
contribution even in this limit. As a consequence, 
the vacuum field does not behave as an adiabatic environment
in the sense of Ref.~\cite{energy}.
In Fig.~2, we plot the diffusion and damping coefficients as functions of 
$\omega_0 t$ for $\omega_0/\Omega=10^{4}.$
They oscillate around their asymptotic values with 
(angular) frequency $\omega_0$ and with an  amplitude of
oscillation that decays in a time  $t \sim 1/\Omega$ ~\cite{foot01}.
The oscillatory 
terms do not contribute to the entropy increase when we average over 
many oscillations.
Hence Eq.~(\ref{deltas}) also holds in this case, although 
the rate of entropy increase is much smaller than in the 
perfect-reflecting limit, since the asymptotic limit of $D_1(t)$ 
is now $D_1=6(\Omega/\omega_0)^2\ln(\omega_0/\Omega)(D_1)_{\rm perf}\ll 
(D_1)_{\rm perf}.$

In order to estimate the decoherence time scale, 
we take, at $t=0,$ the `Schr\"odinger cat' state
$|\psi\rangle = (|\alpha \rangle +|-\alpha \rangle )/\sqrt{2},$
with
$\alpha=iP_0/ \sqrt{2M\hbar \omega_0}.$
The corresponding Wigner function is
\begin{equation}
W=W_m + {1\over \pi\hbar}
\exp \left[-{q^2\over 2(\Delta q)^2}- \label{int}
{2p^2(\Delta q)^2\over \hbar^2} \right] 
\cos(\frac{2 P_0 q}{\hbar}),
\end{equation}
with $\Delta q = \sqrt{\hbar/(2 M \omega_0)},$ and where
$W_m$ corresponds to the statistical mixture
$\rho_m=(1/2)(|\alpha \rangle \langle \alpha| + 
|-\alpha\rangle\langle -\alpha|).$
In phase space, $W_m$ has two peaks along the momentum axis
(at $\pm P_0$). The second term in Eq.~(\ref{int})
originates from the interference between the two state components, and
hence represents the coherence of the cat state.
Since it
oscillates along the $q$ axis in phase space, diffusion in position will
damp the coherence at a maximum rate 
given by
$ -D_1\partial^2_q W/W,$
with, from Eq.~(\ref{int}),
$\partial^2_q W\approx -D_1(2 P_0/\hbar)^2 W.$
After averaging the decoherence rate over a period of
oscillation~\cite{Paz2}, we find that the decoherence time scale
$t_d$ is
\begin{equation}
t_d= {\hbar^2\over  2 P_0^2 D_1}= {\Gamma^{-1}\over
4 |\alpha|^2}.\label{final}
\end{equation}

To clarify the connection between decoherence and the dynamical Casimir
effect, we present a second derivation of Eq.~(\ref{final}), based 
on the concept of entanglement between mirror and field on account
of the generation of photon pairs.
At $t=0,$ 
the
quantum state $|\Psi\rangle$ 
of the complete mirror-plus-field system
is $|\Psi\rangle_0 = |\psi\rangle \otimes |0\rangle$
Instead of tracing over the field operators, we 
follow the evolution of the field state (in the 
interaction picture) to find
$|\Psi\rangle_t = (|\alpha\rangle \otimes |\phi^{+}\rangle_t
+|-\alpha\rangle \otimes |\phi^{-}\rangle_t
)/\sqrt{2},$
where $|\phi^{\pm}\rangle_t$
is computed from
first-order perturbation theory assuming a classical prescribed 
motion:
\begin{equation}
|\phi^{\pm}\rangle_t= B(t) |0\rangle \pm {1\over 2} 
\int_0^{\infty}d\omega_1\int_0^{\infty}d\omega_2 b(\omega_1,\omega_2;t)
|\omega_1,\omega_2\rangle,\label{entangled}
\end{equation}
where 
$$b(\omega_1,\omega_2;t)
={i\over \hbar}\langle \omega_1,\omega_2|{\cal P}|0\rangle
\int_0^tdt' e^{i(\omega_1+\omega_2)t'} {\dot q}(t')$$  
is the two-photon amplitude corresponding to the
mirror's velocity ${\dot q}(t)=-i\sqrt{2\hbar \omega_0/M}\,
\alpha\cos(\omega_0 t)$
associated to the state $|\alpha\rangle,$
whereas $|B(t)|^2$ is determined by the normalization condition 
$\langle \phi^{\pm}| \phi^{\pm} \rangle=1.$
Since the amplitude is proportional to the velocity, it has
an  {\it opposite} sign when associated to $|-\alpha\rangle,$
as shown in Eq.~(\ref{entangled}). 
When $\omega_0 t\gg 1,$ the two-photon probabilities are proportional to 
the time  $t,$ and related to the relaxation rate $\Gamma.$ 
Then, from Eq.~(\ref{entangled}) we derive
$\rho(t)-\rho_m=(1-t/t_d)(\rho(0)-\rho_m),$ with $t_d$ given by~(\ref{final}). 

According to Eq.~(\ref{final}), decoherence is faster than energy
dissipation
by a factor that represents the degree of `macroscopicity'
of the coherent states.
In fact, $|\alpha|^2$ is twice the ratio between the
energy of the coherent state and the zero-point energy of the harmonic oscillator. Therefore, Eq.~(\ref{final}) provides an additional
 illustration of
the meaning of the limit $|\alpha| \gg 1$ as the classical limit of
the quantum harmonic oscillator. Moreover, 
Eq.~(\ref{final}) also shows that
the decoherence rate increases with the distance
beween the two coherent components in
phase space.
 We have confirmed the 
role of coherent states in the understanding of the classical limit 
by showing that they are the pointer states.
Remarkably, classical behavior is
obtained from the mere inclusion of an unavoidable, intrinsically
quantum effect,
the radiation pressure coupling with the quantum vacuum field.

We are grateful to A. Colageracos and G. Barton for correspondence,
and to J. P. Paz, W.
Zurek and S. Haroche for  comments and suggestions. 
DARD thanks UFRJ for its hospitality during his stay, and PAMN 
thanks CNPq and PRONEX for partial financial support.


\begin{references}

\bibitem{Schr} E. Schr\"odinger, Naturwissenschaften {\bf 23}, 807 (1935),
reprinted with English translation in {\it Quantum Theory of Measurement},
edited by J. A. Wheeler and W. H. Zurek (Princeton University Press, Princeton, 1983).

\bibitem{physicstoday} W. H. Zurek, Phys. Today {\bf 44}, No. 10, 36 (1991).

\bibitem{haroche} M. Brune {\it et al.}, Phys. Rev. Lett.
{\bf 77}, 4887 (1996).

\bibitem{luiz} L. Davidovich, M. Brune, J. M. Raimond and S. Haroche,
Phys. Rev. A {\bf  53}, 1295 (1996).

\bibitem{all} A. O. Caldeira and A. J. Leggett, Ann. Phys. (N.Y.)
{\bf 149}, 374 (1983); W. G. Unruh and W. H. Zurek, Phys. Rev. D {\bf 40}, 1071 (1989); B. L. Hu, J. P. Paz and Y. Zhang, Phys. Rev. D {\bf 45}, 2843
(1992).

\bibitem{all-Casimir} P. A. Maia Neto and L. A. S. Machado, Phys. Rev. A
{\bf 54}, 3420 (1996); D. A. R. Dalvit and F. D. Mazzitelli, Phys. Rev. A
{\bf 59}, 3059 (1999); and references therein.

\bibitem{astrid} A. Lambrecht  , M.-T. Jaekel  and S. Reynaud,   Phys. Rev.
Lett. {\bf 77}, 615 (1996).

\bibitem{jr} M. T. Jaekel and S. Reynaud, J. Phys. France {\bf I 3},
 1 (1993).

\bibitem{jr2} M. T. Jaekel and S. Reynaud, Phys. Lett. {\bf A 180},
9 (1993).

\bibitem{barton} G. Barton and A. Calogeracos, Ann. Phys.  (NY) {\bf 238}, 227 (1995);
A. Calogeracos and G. Barton, Ann. Phys. (NY) {\bf 238}, 268 (1995).
Although these references consider a free mirror, the extension to the
harmonic oscillator is straightforward.

\bibitem{ford} L. H. Ford  and A. Vilenkin,   Phys. Rev. D {\bf  25}, 2569 (1982).

\bibitem{Paz} W. H. Zurek, S. Habib and J. P. Paz,
Phys. Rev. Lett. {\bf 70}, 1187 (1993).

\bibitem{energy} J. P. Paz and W. H. Zurek,
Phys. Rev. Lett. {\bf 82}, 5181 (1999).

\bibitem{foot01} 
If a stronger high-frequency cut--off is introduced in our model, 
so as to render the correlation function 
$\sigma(0)=\int_0^{\infty} d\omega\Xi(\omega)/\pi$ finite, 
then  it would follow from Eq.~(\ref{D1})  that $D_1\approx 
(\sigma(0)/2 M^2 \omega_0) \sin(\omega_0 t)$
when $\omega_0$ is much larger than the
frequency cut--off. In this case, the vacuum field behaves as 
an adiabatic environment coupled linearly to the harmonic oscillator,
and, as discussed in Ref.~\cite{energy}, no decoherence takes place.

\bibitem{Paz2} The decoherence rate decreases as the 
two coherent components of $|\psi\rangle$ rotate away from their 
initial positions in phase space, so that the average 
rate is one-half of the maximum value.
When the interaction is linear in the position rather than in the momentum 
operator, 
as in the usual QBM models [J. P. Paz, S. Habib and
W. H. Zurek, Phys. Rev. D {\bf 47}, 488 (1993)], 
the opposite applies,  and then the
maximum rate occurs when the cat state is along the position axis.



\end{references}
\end{document}